\documentclass[twocolumn,showpacs,preprintnumbers,amsmath,amssymb]{revtex4}

\usepackage{graphicx}
\usepackage{dcolumn}
\usepackage{bm}
\begin{document}

\preprint{VERSION 0.4}

\title{Theory of Sodium Ordering in Na$_x$CoO$_2$}

\author{Peihong Zhang$^{1,2}$, Rodrigo B. Capaz$^{1,2,3}$, Marvin L. Cohen$^{1,2}$, and Steven G. Louie$^{1,2}$}

\affiliation{$^1$ Department of Physics, University of California at Berkeley, Berkeley, CA 94720 \\
$^2$ Materials Science Division, Lawrence Berkeley National Laboratory, Berkeley, CA 94720 \\
$^3$ Instituto de F\'\i sica, Universidade Federal do Rio de Janeiro, Caixa Postal 68528, Rio de Janeiro, RJ 21941-972, Brazil
}

\date{\today}

\begin{abstract}
The ordering of Na ions in Na$_x$CoO$_2$ is investigated systematically 
by combining detailed density functional theory (DFT) studies with model
calculations. Various ground state ordering patterns are identified,
and they are in excellent agreement with avaliable experimental results. Our results 
suggest that the primary driving force for the Na ordering is the screened
Coulomb interaction among Na ions. Possible effects of the Na ordering
on the electronic structure of the CoO$_2$ layer are discussed. We propose
that the nonexistence of a charge ordered insulating state at $x = 2/3$ is 
due to the lack of a commensurate Na ordering pattern, whereas an extremely
stable Na ordering at $x = 0.5$ enhances the charge ordering tendency, resulting
in an insulating state as observed experimentally.
\end{abstract}

\maketitle


The recently renewed research interest in Na$_x$CoO$_2$ since the discovery
of superconductivity \cite{supercond} in the hydrated materials has revealed a 
range of interesting and intriguing properties of this system. One of the 
most interesting discoveries is the determination of its phase diagram as the 
doping level $x$ is varied \cite{Foo04}: Two metallic phases at low and high dopings
are separated by an insulating state at $x=0.5$. Along with this 
insulating state there is strong evidence of an ordered Na layer and commensurate 
charge orderings in the CoO$_2$ layer. This raises the possibility of a 
subtle interplay between the Na ordering and the charge ordering in the 
CoO$_2$ layer\cite{Foo04,huang2}. At certain Na compositions, this interplay might have
profound effects on the electronic or magnetic properties of the CoO$_2$ layer
if a particularly stable Na ordering pattern exists and whether a commensurate charge 
ordering is allowed. Therefore, a unified theory of the Na ordering mechanism may be 
critical for a better understanding of the electronic structure in this system.
The observed Na ordering pattern \cite{Foo04,zandbergen} at $x=0.5$ 
agrees well with our theoretical prediction \cite{zhang}, and this motivates us to 
extend our investigation on Na ordering at other 
compositions studied by recent electron diffraction experiments \cite{zandbergen}. 
In this paper, we identify ground state Na ordering patterns at various
compositions and show that the Na ordering is primarily driven by the intra-plane 
screened electrostatic interactions. The details of the electronic structure
of the CoO$_2$ plane, on the other hand, have minimal effects. 

The compound Na$_x$CoO$_2$ assumes a layered structure which consists of alternating 
triangular CoO$_2$ and Na planes. Although there have been 
reports \cite{Delmas81,Balsys96,Ono02,Sugiyama04} of several stacking patterns,
we shall restrict our discussion to the so-called $\gamma$ phase, which is usually 
observed for low to intermediate Na compositions. There are two distinct Na sites 
within a given plane, denoted as Na(1) and Na(2) (Wyckoff indices $2b$ and $2d$). 
The Na(1) site, being directly between two Co ions, is slightly higher 
in energy than the Na(2) site\cite{zhang}. Therefore, 
Na ions will normally prefer occupying the Na(2) sites.

We calculate the total energy of 82 ordered structures of Na$_x$CoO$_2$ 
using first principles techniques. The choice of structures for the ground 
state search was guided by the MAPS code \cite{maps}. Our calculations are 
based on density-functional theory (DFT) \cite{hk,ks} within the local density 
approximation (LDA) with the Ceperly-Alder exchange-correlation 
functional \cite{ca,pz}. {\it Ab initio} Troullier-Martins pseudopotentials \cite{tm} 
are used. Calculations are performed using the SIESTA code \cite{siesta1,siesta2}, 
which expands the Kohn-Sham wavefunctions in a 
linear combination of atomic orbitals (LCAO). A double-zeta plus polarization
(DZP) basis is used. The energy cutoff for the charge density grid is 1000 Ry. 
The irreducible Brillouin Zone is sampled using Monkhorst-Pack grids \cite{mp} 
with a density of k-points equivalent to $12\times12\times2$ for
a primitive unit cell. Atomic positions are relaxed until forces are smaller than 
0.005 eV/\AA. We keep the lattice constants fixed at $a=2.82{ }$\AA{ }and $c=10.89{ }$\AA.
Although the lattice constants of Na$_x$CoO$_2$ will vary slightly
with Na concentration, our results are not sensitive
to this small variation.

\begin{figure}
\includegraphics[width=3.2in]{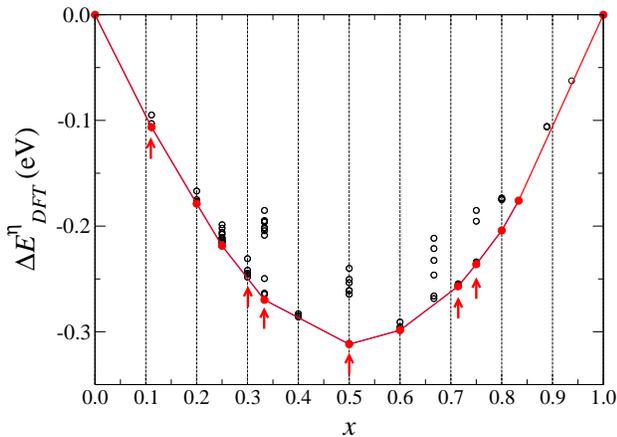}
\caption{DFT formation energies for 82 ordered structures of Na$_x$CoO$_2$. The
red dots represent the ground-state structures, joined together by the convex hull
(red line). The arrows indicate the experimentally observed ground states \cite{zandbergen}.}
\label{fig.1}
\end{figure}

\begin{figure}
\includegraphics[width=3.1in]{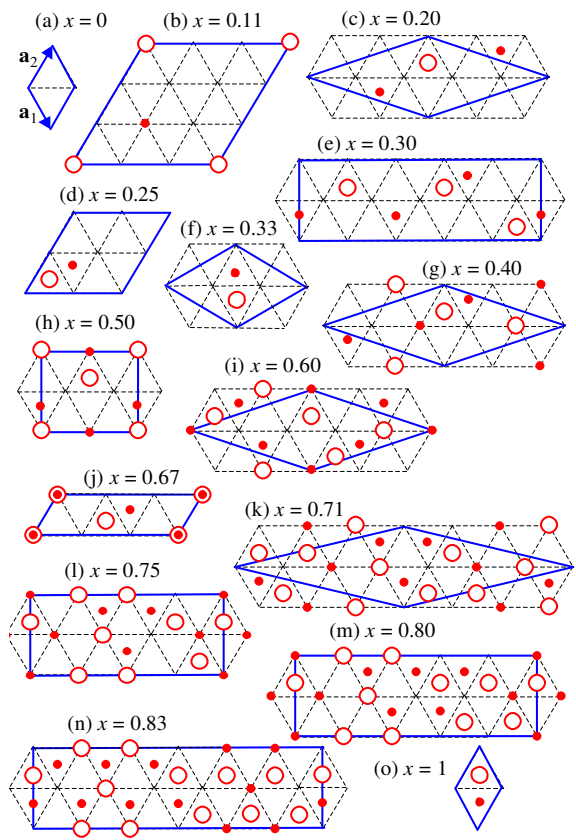}
\caption{Relevant ordered structures of Na$_x$CoO$_2$. The dashed lines represent the
projected triangular lattice of Co atoms. Large red open circles and small red dots
represent projected Na positions in different planes, at $z=0$ and $z=0.5c$, respectively. 
The Na(1) sites fall on top of the triangular lattice sites and Na(2) sites fall in the 
center of the triangles. Blue lines indicate the minimum unit cell in each case. 
A detailed discussion of each structure is presented in the text.}
\label{fig.2}
\end{figure}

The energetics of Na ordering in Na$_x$CoO$_2$ can be studied using the 
well-stablished formalism for binary alloys \cite{defontaine}. The 
fundamental quantity that determines the stability of a given arrangement 
$\eta$ of Na ions and vacancies with Na composition $x$ is the formation 
energy $\Delta E^\eta$:
\begin{equation}
\label{eq.1}
\Delta E_{DFT}^\eta = E^\eta - xE^{NaCoO_2} - (1-x)E^{CoO_2},
\end{equation}
where $E^\eta$, $E^{NaCoO_2}$ and $E^{CoO_2}$ are the total energies {\it per site} of the
structure $\eta$, and of the pure compounds NaCoO$_2$ and CoO$_2$, respectively, calculated
within DFT. Fig. \ref{fig.1}
shows the calculated formation energies of all 82 structures (open dots). We highlight 
the ground state structures (red dots) joined together by the hull-shaped curve
(red line). Structures with
formation energy above the convex hull are unstable against phase 
separation into the two ground 
states with nearby compositions. Red arrows indicate the structures proposed by Zandbergen
{\it et al.} \cite{zandbergen} (at $x=0.11$, $x=0.30$, $x=0.33$, $x=0.50$, $x=0.71$ and $x=0.75$)
based on electron diffraction experiments. The agreement between theory and experiment 
is remarkable in that all experimentally proposed structures are indeed
ground states within DFT.
Although the structure at $x=0.30$ appears to be 
an exception (it is not a ground state within DFT), it is too close to the
convex hull to be neglected as such, given the uncertainties in the calculation. 
Moreover, we find additional 
ground states at $x=0.20$, $x=0.25$, $x=0.60$, $x=0.80$ and $x=0.83$. A structure at $x=0.40$ is  
also too close to the convex hull to be ruled out. These structures have not 
been observed experimentally, perhaps 
because samples at these particular compositions were not analyzed. An ordered structure at $x=0.64$
also appears experimentally \cite{zandbergen}. According to our results, at this composition 
the sample should be phase-separated between the $x=0.60$ and the $x=0.71$ ground states. 
All ground state structures
and the near ground states at $x=0.30$ and $x=0.40$ are depicted in Fig.\ref{fig.2}. We also
display the lowest energy structure at $x=0.67$ (not a ground state), for discussion purposes.

Before discussing in detail the various ordered structures, let us first address 
the driving force for Na ordering in this system. The simplest explanation would 
be electrostatics since the Na atoms lose one electron each to the CoO$_2$ network,
they arrange themselves so as to minimize electrostatic energy. We perform model calculations 
to test this hypothesis, in which the Ewald (ionic) energies of the different arrangements
of Na$^{+1}$ ions are computed. In these calculations, we assume that the electronic 
charge is spread over the CoO$_2$ planes neighboring the Na-vacancy layer, which is  
consistent with our previous calculations \cite{zhang}. Also, 
the chemical difference between Na(1) and Na(2) sites is simulated by adding an energy 
$\epsilon_1$ per Na ion to the Ewald energies for a Na(1) site.
Therefore, the total energy per unit cell for this electrostatic model is defined as:
\begin{equation}
\label{eq.2}
E_{model}^{\eta} = E_{Ewald}^{\eta} + N_1^{\eta}\epsilon_1,
\end{equation}
where $N_1^{\eta}$ is the number of Na atoms in Na(1) sites for the structure $\eta$.

\begin{figure}
\includegraphics[width=3.2in]{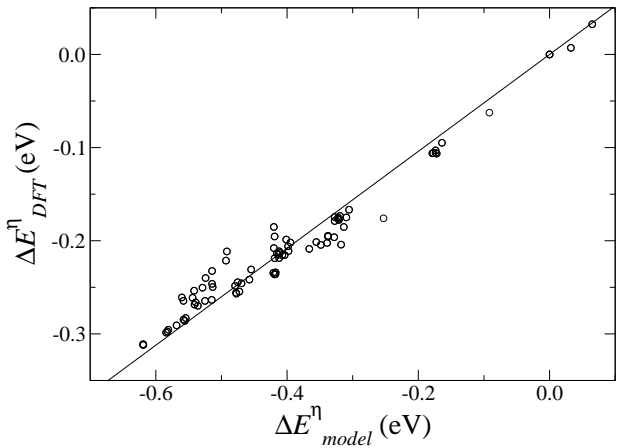}
\caption{DFT versus electrostatic model formation energies.}
\label{fig.3}
\end{figure}

Figure \ref{fig.3} shows, for each structure, $\Delta E^{\eta}_{DFT}$ (Eq.(\ref{eq.1})) 
versus the calculated formation energy using the above model 
($\Delta E^{\eta}_{model}$). We clearly see that these two quantities are proportional:
$\Delta E^{\eta}_{DFT} = \kappa \Delta E^{\eta}_{model}$.
The best fit (rms deviation of 0.016 eV) is obtained for $\kappa = 0.514$ and 
an on-site energy difference between Na(1) and Na(2) sites $\kappa\epsilon_1 = 67$ meV.
A value of $\kappa$ smaller than 1 is expected because of screening and partial ionization 
effects, which are absent from the model calculations and naturally included in the DFT calculations. 
Therefore, DFT formation energies 
can be described with good accuracy by a model of screened electrostatic
interactions, showing unambiguously that {\it the dominant driving force for Na ordering in 
Na$_x$CoO$_2$ is electrostatic}. 

We now discuss the geometries of the structures shown in Fig. \ref{fig.2}. 
In the low concentration regime, we find a ground state at $x=0.11$ (1/9) (Fig. \ref{fig.2}(b)), 
corresponding to an in-plane arrangement
of Na ``impurities'' in a $3\times3$ triangular lattice. Surprisingly, the lowest 
energy structure corresponds to placing both Na atoms (one in each plane) in the less favorable Na(1) 
sites. This structure prevails with respect to other arrangements because it minimizes the inter-plane 
electrostatic energy: The Na atoms can be as far as possible from each other. Interestingly, this
arrangement corresponds to placing each Na atom in the middle of the projections of the three Na atoms
of neighboring planes, and it is consistent with  the experimental observation of a 
$\sqrt{3}\times\sqrt{3}$ superstructure in the diffraction patterns of Ref. \cite{zandbergen} 
if a random stacking of such arrangements occurs.

At $x=0.20$, the ground state has a $\sqrt{7}\times\sqrt{7}$ supercell, shown in 
Fig.\ \ref{fig.2}(c), with both Na ions in Na(2) sites. Periodic repetition of 
this cell leads to an arrangement of Na ions along rows perpendicular to the [110] 
direction (the horizontal direction in Fig. 2). The distance between Na ions in a 
row is $\sqrt{3}a$. Such pattern of 
rows have been experimentally observed for different compositions and have been 
proposed to be the guiding principle for Na ordering \cite{zandbergen}.
Indeed, we observe this pattern in all other Na compositions except
$x = 1/9$, 1/4, and 2/3 (not a ground state). 
The same pattern of rows is observed in the lowest energy structures calculated using 
the simple electrostatic model. Therefore, {\it the rows arise naturally from the 
minimization of electrostatic energy, combined with the constraint that only 
discrete lattice sites can be occupied by Na ions}.


Fig. \ref{fig.2}(d) shows the ground state for $x=0.25$, a $2\times2$ triangular 
lattice. In this case, both Na ions occupy the more preferable Na(2) sites. 
In Fig. \ref{fig.2}(e), the lowest energy structure for $x=0.30$ is shown.
It has precisely the same in-plane arrangement proposed by 
Zandbergen {\it et al.}\cite{zandbergen}. Again, it corresponds 
to rows of Na ions perpendicular to [110] and all ions occupy Na(2) sites. 
As discussed above, it is not a ground state in our calculations (by merely 1 meV/site).

At $x=0.33$ (1/3), the arrangement is again a triangular lattice ($\sqrt{3}\times\sqrt{3}$), 
as shown in Fig. \ref{fig.2}(f). This structure is exactly the same as proposed by 
Zandbergen {\it et al.} \cite{zandbergen}. The $x=1/3$ concentration is of
particular interest due to the observed superconductivity in the hydrated
system at $x\sim0.3$. Although the role water plays is still unknown, the screening 
of the Na potential by water molecules might suppress the tendency of developing a commensurate 
charge ordering pattern in the CoO$_2$ layer, which may then lead to a more 
homogeneous electronic system and favor a superconducting state over other 
competing phases at low temperature. Possible charge disproportionation  
and gap opening at this doping level in the unhydrated system has been studied 
previously \cite{Lee04_2}.  However, so far such an insulating state has not been observed. 


Fig. \ref{fig.2}(g) shows the lowest energy arrangement for $x=0.40$ (not a ground state, 
by 0.6 meV/site). It also follows the rows pattern, and the ratio between
Na(1) and Na(2) occupancies is 1. 

In Fig. \ref{fig.2}(h), the ground state for $x=0.50$ is shown, which is 
a $\sqrt{3}\times2$ supercell. The arrangement also follows the pattern of 
rows and it has also been predicted theoretically \cite{zhang} and 
observed experimentally \cite{zandbergen}. This structure is particularly stable since
all other in-plane arrangements with a similar unit cell at this composition are 
at least 25 meV/site (or 200 meV/cell) higher in energy.
Note again that the ratio between Na(1) and Na(2) occupancies is 1. This ratio 
tends to be higher for structures with intermediate composition, 
since the on-site chemical energy difference between Na(1) and Na(2) sites becomes 
less important than arranging the Na ions as far as possible to minimize electrostatic energy. 
The existence of a particularly stable Na ordering pattern, together with
the observed insulating state at this composition \cite{Foo04}, strongly
indicate an interplay between the Na 
ordering and the charge ordering in the CoO$_2$ layer in this system. 
Our results suggest that, if such an interplay exists, it is more likely
that an ordered Na potential drives the charge ordering or enhances 
this tendency. Studies of charge ordering in this system with the presence of
a commensurate Na ordering will be reported in a separate paper.

At $x=0.60$, the ground state structure is again a $\sqrt{7}\times\sqrt{7}$ supercell, 
shown in Fig. \ref{fig.2}(i) with a pattern of rows and a Na(1)/Na(2) occupancy ratio of 1/2. 
The composition $x=0.67$ (2/3), like $x=1/3$,  has been the subject of many discussions
due to its apparent proximity to charge ordering in a triangular
lattice\cite{Lee04_2,Motrunich04,Mukhamedshin04}. The lowest energy structure 
corresponds to $3\times1$ supercell, shown in Fig. \ref{fig.2}(j). 
This structure is 
clearly lower in energy than any honeycomb arrangement of Na ions. It is obvious
that a honeycomb arrangement would be more compatible with the proposed charge ordering 
pattern\cite{Lee04_2,Motrunich04,Mukhamedshin04}. Therefore, we propose
that the nonexistence of a charge-ordered insulating state at this 
composition is due to the lack of a stable and commensurate Na ordering.

The ground state structure at $x=0.71$ (5/7) is shown in Fig. \ref{fig.2}(k). It is 
impressive that such a complex arrangement with a large supercell ($\sqrt{13}\times\sqrt{13}$) 
corresponds to a ground state structure, as seen experimentally \cite{zandbergen}.
For this structure we have a Na(1)/Na(2) occupancy ratio of 2/3 and once more the rows 
pattern is followed. In the high $x$ limit, this pattern becomes simpler to analyze: 
Low energy structures are composed of rows of Na(2) ions intercalated by 
rows of vacancies and Na(1) ions. For instance, for the $x=0.71$ structure the 
repetition is 3$\times$Na(2)-vac-2$\times$Na(1)-vac. 

For $x=0.75$ we find again the same ground state as proposed by experiments \cite{zandbergen}, 
a $\sqrt{3}\times4$ supercell shown in Fig. \ref{fig.2}(l). In this case, the pattern of rows 
is 3$\times$Na(2)-vac-3$\times$Na(1)-vac, i.e., there are equal numbers 
of Na(2) and Na(1) ions. This structure is lower in energy than that proposed
by Shi {\it et al.}\cite{Shi03}. Measurements of local distortions of Na(2)O$_6$ polyhedra 
at $x=0.75$ have been recently performed by neutron diffraction \cite{huang1}. Apparently, 
Na(2) ions move to off-center positions, resulting in two long (2.56 \AA) and four short (2.32 \AA)
Na-O bonds. Our calculated Na-O bond lengths for this structure are 2.57 \AA{ } and 2.29 \AA, 
in excellent agreement with experiment. 

At $x=0.80$ and $x=0.83$ (5/6) (Figs. \ref{fig.2} (m) and (n), respectively) we find ground-state
structures that belong to the same ``family'' as the $x=0.75$ 
structure, i.e., they follow similar row patterns with $n\times$Na(2)-vac-3$\times$Na(1)-vac 
rows in a $\sqrt{3}\times (n+5)/2$ cell, with $n=3$ for $x=0.75$, $n=5$ for $x=0.80$ and $n=7$ for
$x=0.83$. Therefore, it is very likely that beyond $x=0.75$ one can construct 
infinitely many ground states (polytipoids) by simply intercalating odd
numbers of Na(2) rows with the sequence of rows -vac-3$\times$Na(1)-vac-. 
In fact, if one takes into account entropic considerations, it is plausible that the 
thermodynamically stable configuration 
for $x>0.75$ will consist of disordered arrangements of such rows. This result may be related to 
the order-disorder transition reported at $x=0.75$ from neutron diffraction experiments \cite{huang1,huang2}. 
Finally, for $x=1.0$ all Na(2) sites are occupied (Fig. \ref{fig.2}(o)).

In conclusion, we have carried out detailed DFT and model studies on the Na 
ordering mechanism in Na$_x$CoO$_2$. The ordering pattern is non-trivial and 
is sensitive to the Na concentration. The theoretically determined ordering patterns at 
various Na compositions agree well with those observed in recent electron 
diffraction experiments\cite{zandbergen}. In addition, we identify several ground states
that have not yet been observed. Our results indicate that the
primary driving force for the Na ordering is the screened electrostatic
interactions. Detailed electronic structure on the CoO$_2$ layer,
on the other hand, plays no important role on the ordering of
Na ions since the coupling between the CoO$_2$ plane and the Na plane 
is much weaker than the intra-plane Na-Na interactions. The converse
effects, however, could be significant since the charge and magnetic orderings 
on the CoO$_2$ plane happen at extremely low energy scale\cite{zhang} and
may be vulnerable to external perturbations.
Therefore, it is likely that the particularly stable Na 
ordering structure at $x = 0.5$ enhances the charge ordering tendency 
in the CoO$_2$ layer, resulting in a charge ordered insulating state 
at low temperature as observed experimentally\cite{Foo04}. 
Conversely, the nonexistence of such an insulating state at $x =
2/3$ may be due to the absence of an ordered honeycomb arrangement of Na
ions at this composition.

We acknowledge useful discussions with A. van de Walle, H. W. Zandbergen, and Q. Huang. 
This work was partially supported by National Science Foundation Grant No. 
DMR04-39768 and by the Director, Office of Science, Office of Basic 
Energy Sciences, Division of Materials Sciences and Engineering, U.S. 
Department of Energy under Contract No. DE-AC03-76SF00098.
RBC acknowledges financial support from the 
John Simon Guggenheim Memorial Foundation and Brazilian funding agencies 
CNPq, CAPES, FAPERJ, Instituto de Nanoci{\^e}ncias, FUJB-UFRJ and PRONEX-MCT. 
Computational resources were provided by NPACI and NERSC.

\bibliography{ord-0.4}

\begin{thebibliography}{24}
\expandafter\ifx\csname natexlab\endcsname\relax\def\natexlab#1{#1}\fi
\expandafter\ifx\csname bibnamefont\endcsname\relax
  \def\bibnamefont#1{#1}\fi
\expandafter\ifx\csname bibfnamefont\endcsname\relax
  \def\bibfnamefont#1{#1}\fi
\expandafter\ifx\csname citenamefont\endcsname\relax
  \def\citenamefont#1{#1}\fi
\expandafter\ifx\csname url\endcsname\relax
  \def\url#1{\texttt{#1}}\fi
\expandafter\ifx\csname urlprefix\endcsname\relax\def\urlprefix{URL }\fi
\providecommand{\bibinfo}[2]{#2}
\providecommand{\eprint}[2][]{\url{#2}}

\bibitem[{\citenamefont{Takada et~al.}(2003)\citenamefont{Takada, Sakural,
  Takayama-Muromachi, Izumi, Dilanian, and Sasaki}}]{supercond}
\bibinfo{author}{\bibfnamefont{K.}~\bibnamefont{Takada}},
  \bibinfo{author}{\bibfnamefont{H.}~\bibnamefont{Sakural}},
  \bibinfo{author}{\bibfnamefont{E.}~\bibnamefont{Takayama-Muromachi}},
  \bibinfo{author}{\bibfnamefont{F.}~\bibnamefont{Izumi}},
  \bibinfo{author}{\bibfnamefont{R.~A.} \bibnamefont{Dilanian}},
  \bibnamefont{and} \bibinfo{author}{\bibfnamefont{T.}~\bibnamefont{Sasaki}},
  \bibinfo{journal}{Nature (London)} \textbf{\bibinfo{volume}{422}},
  \bibinfo{pages}{53} (\bibinfo{year}{2003}).

\bibitem[{\citenamefont{Foo et~al.}(2004)\citenamefont{Foo, Wang, Watauchi,
  Zandbergen, He, Cava, and Ong}}]{Foo04}
\bibinfo{author}{\bibfnamefont{M.}~\bibnamefont{Foo}},
  \bibinfo{author}{\bibfnamefont{Y.}~\bibnamefont{Wang}},
  \bibinfo{author}{\bibfnamefont{S.}~\bibnamefont{Watauchi}},
  \bibinfo{author}{\bibfnamefont{H.~W.} \bibnamefont{Zandbergen}},
  \bibinfo{author}{\bibfnamefont{T.}~\bibnamefont{He}},
  \bibinfo{author}{\bibfnamefont{R.~J.} \bibnamefont{Cava}}, \bibnamefont{and}
  \bibinfo{author}{\bibfnamefont{N.~P.} \bibnamefont{Ong}},
  \bibinfo{journal}{Phys.\ Rev.\ Lett.} \textbf{\bibinfo{volume}{92}},
  \bibinfo{pages}{247001} (\bibinfo{year}{2004}).

\bibitem[{\citenamefont{Huang et~al.}(2004{\natexlab{a}})\citenamefont{Huang,
  Foo, Pascal, Lynn, Toby, He, Zandbergen, and Cava}}]{huang2}
\bibinfo{author}{\bibfnamefont{Q.}~\bibnamefont{Huang}},
  \bibinfo{author}{\bibfnamefont{M.~L.} \bibnamefont{Foo}},
  \bibinfo{author}{\bibfnamefont{R.~A.} \bibnamefont{Pascal}},
  \bibinfo{author}{\bibfnamefont{J.~W.} \bibnamefont{Lynn}},
  \bibinfo{author}{\bibfnamefont{B.~H.} \bibnamefont{Toby}},
  \bibinfo{author}{\bibfnamefont{T.}~\bibnamefont{He}},
  \bibinfo{author}{\bibfnamefont{H.}~\bibnamefont{Zandbergen}},
  \bibnamefont{and} \bibinfo{author}{\bibfnamefont{R.~J.} \bibnamefont{Cava}},
  \bibinfo{journal}{Phys. Rev. B} \textbf{\bibinfo{volume}{70}},
  \bibinfo{pages}{184110} (\bibinfo{year}{2004}{\natexlab{a}}).

\bibitem[{\citenamefont{Zandbergen et~al.}(2004)\citenamefont{Zandbergen, Foo,
  Xu, Kumar, and Cava}}]{zandbergen}
\bibinfo{author}{\bibfnamefont{H.~W.} \bibnamefont{Zandbergen}},
  \bibinfo{author}{\bibfnamefont{M.}~\bibnamefont{Foo}},
  \bibinfo{author}{\bibfnamefont{Q.}~\bibnamefont{Xu}},
  \bibinfo{author}{\bibfnamefont{V.}~\bibnamefont{Kumar}}, \bibnamefont{and}
  \bibinfo{author}{\bibfnamefont{R.~J.} \bibnamefont{Cava}},
  \bibinfo{journal}{Phys.\ Rev.\ B} \textbf{\bibinfo{volume}{70}},
  \bibinfo{pages}{024101} (\bibinfo{year}{2004}).

\bibitem[{\citenamefont{Zhang et~al.}(2004)\citenamefont{Zhang, Luo, Crespi,
  Cohen, and Louie}}]{zhang}
\bibinfo{author}{\bibfnamefont{P.}~\bibnamefont{Zhang}},
  \bibinfo{author}{\bibfnamefont{W.}~\bibnamefont{Luo}},
  \bibinfo{author}{\bibfnamefont{V.~H.} \bibnamefont{Crespi}},
  \bibinfo{author}{\bibfnamefont{M.~L.} \bibnamefont{Cohen}}, \bibnamefont{and}
  \bibinfo{author}{\bibfnamefont{S.~G.} \bibnamefont{Louie}},
  \bibinfo{journal}{Phys.\ Rev.\ B} \textbf{\bibinfo{volume}{70}},
  \bibinfo{pages}{085108} (\bibinfo{year}{2004}).

\bibitem[{\citenamefont{Delmas et~al.}(1981)\citenamefont{Delmas, Braconnier,
  Fouassier, and Hagenmuller}}]{Delmas81}
\bibinfo{author}{\bibfnamefont{D.}~\bibnamefont{Delmas}},
  \bibinfo{author}{\bibfnamefont{J.~J.} \bibnamefont{Braconnier}},
  \bibinfo{author}{\bibfnamefont{C.}~\bibnamefont{Fouassier}},
  \bibnamefont{and}
  \bibinfo{author}{\bibfnamefont{P.}~\bibnamefont{Hagenmuller}},
  \bibinfo{journal}{Solid State Ionics} \textbf{\bibinfo{volume}{3-4}},
  \bibinfo{pages}{165} (\bibinfo{year}{1981}).

\bibitem[{\citenamefont{Balsys and Davis}(1996)}]{Balsys96}
\bibinfo{author}{\bibfnamefont{R.~J.} \bibnamefont{Balsys}} \bibnamefont{and}
  \bibinfo{author}{\bibfnamefont{R.~L.} \bibnamefont{Davis}},
  \bibinfo{journal}{Solid State Ionics} \textbf{\bibinfo{volume}{93}},
  \bibinfo{pages}{279} (\bibinfo{year}{1996}).

\bibitem[{\citenamefont{Ono et~al.}(2002)\citenamefont{Ono, Ishikawa, Miyazaki,
  Ishii, Morii, and Kajitani}}]{Ono02}
\bibinfo{author}{\bibfnamefont{Y.}~\bibnamefont{Ono}},
  \bibinfo{author}{\bibfnamefont{R.}~\bibnamefont{Ishikawa}},
  \bibinfo{author}{\bibfnamefont{Y.}~\bibnamefont{Miyazaki}},
  \bibinfo{author}{\bibfnamefont{Y.}~\bibnamefont{Ishii}},
  \bibinfo{author}{\bibfnamefont{Y.}~\bibnamefont{Morii}}, \bibnamefont{and}
  \bibinfo{author}{\bibfnamefont{T.}~\bibnamefont{Kajitani}},
  \bibinfo{journal}{J. Solid State Chem.} \textbf{\bibinfo{volume}{166}},
  \bibinfo{pages}{177} (\bibinfo{year}{2002}).

\bibitem[{\citenamefont{Sugiyama et~al.}(2004)\citenamefont{Sugiyama, Brewer,
  Ansaldo, Hitti, Mikami, Mori, and Sasaki}}]{Sugiyama04}
\bibinfo{author}{\bibfnamefont{J.}~\bibnamefont{Sugiyama}},
  \bibinfo{author}{\bibfnamefont{J.~H.} \bibnamefont{Brewer}},
  \bibinfo{author}{\bibfnamefont{E.~J.} \bibnamefont{Ansaldo}},
  \bibinfo{author}{\bibfnamefont{B.}~\bibnamefont{Hitti}},
  \bibinfo{author}{\bibfnamefont{M.}~\bibnamefont{Mikami}},
  \bibinfo{author}{\bibfnamefont{Y.}~\bibnamefont{Mori}}, \bibnamefont{and}
  \bibinfo{author}{\bibfnamefont{T.}~\bibnamefont{Sasaki}},
  \bibinfo{journal}{Phys.\ Rev.\ B} \textbf{\bibinfo{volume}{69}},
  \bibinfo{pages}{214423} (\bibinfo{year}{2004}).

\bibitem[{\citenamefont{van~de Walle and Ceder}(2002)}]{maps}
\bibinfo{author}{\bibfnamefont{A.}~\bibnamefont{van~de Walle}}
  \bibnamefont{and} \bibinfo{author}{\bibfnamefont{G.}~\bibnamefont{Ceder}},
  \bibinfo{journal}{J. Phase Equilibria} \textbf{\bibinfo{volume}{23}},
  \bibinfo{pages}{248} (\bibinfo{year}{2002}).

\bibitem[{\citenamefont{Hohenberg and Kohn}(1964)}]{hk}
\bibinfo{author}{\bibfnamefont{P.}~\bibnamefont{Hohenberg}} \bibnamefont{and}
  \bibinfo{author}{\bibfnamefont{W.}~\bibnamefont{Kohn}},
  \bibinfo{journal}{Phys.\ Rev.} \textbf{\bibinfo{volume}{136}},
  \bibinfo{pages}{B864} (\bibinfo{year}{1964}).

\bibitem[{\citenamefont{Kohn and Sham}(1965)}]{ks}
\bibinfo{author}{\bibfnamefont{W.}~\bibnamefont{Kohn}} \bibnamefont{and}
  \bibinfo{author}{\bibfnamefont{L.}~\bibnamefont{Sham}},
  \bibinfo{journal}{Phys.\ Rev.} \textbf{\bibinfo{volume}{140}},
  \bibinfo{pages}{A1133} (\bibinfo{year}{1965}).

\bibitem[{\citenamefont{Ceperley and Alder}(1980)}]{ca}
\bibinfo{author}{\bibfnamefont{D.~M.} \bibnamefont{Ceperley}} \bibnamefont{and}
  \bibinfo{author}{\bibfnamefont{B.~J.} \bibnamefont{Alder}},
  \bibinfo{journal}{Phys.\ Rev. Lett.} \textbf{\bibinfo{volume}{45}},
  \bibinfo{pages}{566} (\bibinfo{year}{1980}).

\bibitem[{\citenamefont{Perdew and Zunger}(1981)}]{pz}
\bibinfo{author}{\bibfnamefont{J.~P.} \bibnamefont{Perdew}} \bibnamefont{and}
  \bibinfo{author}{\bibfnamefont{A.}~\bibnamefont{Zunger}},
  \bibinfo{journal}{Phys.\ Rev. B} \textbf{\bibinfo{volume}{23}},
  \bibinfo{pages}{5048} (\bibinfo{year}{1981}).

\bibitem[{\citenamefont{Troullier and Martins}(1991)}]{tm}
\bibinfo{author}{\bibfnamefont{N.}~\bibnamefont{Troullier}} \bibnamefont{and}
  \bibinfo{author}{\bibfnamefont{J.~L.} \bibnamefont{Martins}},
  \bibinfo{journal}{Phys.\ Rev. B} \textbf{\bibinfo{volume}{43}},
  \bibinfo{pages}{1993} (\bibinfo{year}{1991}).

\bibitem[{\citenamefont{Ordej\'on et~al.}(1996)\citenamefont{Ordej\'on,
  Artacho, and Soler}}]{siesta1}
\bibinfo{author}{\bibfnamefont{P.}~\bibnamefont{Ordej\'on}},
  \bibinfo{author}{\bibfnamefont{E.}~\bibnamefont{Artacho}}, \bibnamefont{and}
  \bibinfo{author}{\bibfnamefont{J.~M.} \bibnamefont{Soler}},
  \bibinfo{journal}{Phys. Rev. B} \textbf{\bibinfo{volume}{53}},
  \bibinfo{pages}{R10441} (\bibinfo{year}{1996}).

\bibitem[{\citenamefont{Soler et~al.}(2002)\citenamefont{Soler, Artacho, Gale,
  Garc\'{\i}a, Junquera, Ordej\'on, and S\'anchez-Portal}}]{siesta2}
\bibinfo{author}{\bibfnamefont{J.~M.} \bibnamefont{Soler}},
  \bibinfo{author}{\bibfnamefont{E.}~\bibnamefont{Artacho}},
  \bibinfo{author}{\bibfnamefont{J.~D.} \bibnamefont{Gale}},
  \bibinfo{author}{\bibfnamefont{A.}~\bibnamefont{Garc\'{\i}a}},
  \bibinfo{author}{\bibfnamefont{J.}~\bibnamefont{Junquera}},
  \bibinfo{author}{\bibfnamefont{P.}~\bibnamefont{Ordej\'on}},
  \bibnamefont{and}
  \bibinfo{author}{\bibfnamefont{D.}~\bibnamefont{S\'anchez-Portal}},
  \bibinfo{journal}{J. Phys.: Condens. Matter} \textbf{\bibinfo{volume}{14}},
  \bibinfo{pages}{2745} (\bibinfo{year}{2002}).

\bibitem[{\citenamefont{Monkhorst and Pack}(1976)}]{mp}
\bibinfo{author}{\bibfnamefont{H.~J.} \bibnamefont{Monkhorst}}
  \bibnamefont{and} \bibinfo{author}{\bibfnamefont{J.~D.} \bibnamefont{Pack}},
  \bibinfo{journal}{Phys. Rev. B} \textbf{\bibinfo{volume}{13}},
  \bibinfo{pages}{5188} (\bibinfo{year}{1976}).

\bibitem[{\citenamefont{de~Fontaine}(1994)}]{defontaine}
\bibinfo{author}{\bibfnamefont{D.}~\bibnamefont{de~Fontaine}}, in
  \emph{\bibinfo{booktitle}{Solid State Physics}}, edited by
  \bibinfo{editor}{\bibfnamefont{H.}~\bibnamefont{Ehrenheich}}
  \bibnamefont{and} \bibinfo{editor}{\bibfnamefont{D.}~\bibnamefont{Turnbull}}
  (\bibinfo{publisher}{Academic, New York}, \bibinfo{year}{1994}),
  p.~\bibinfo{pages}{33}.

\bibitem[{\citenamefont{Lee et~al.}(2004)\citenamefont{Lee, Kune\v{s}, and
  Pickett}}]{Lee04_2}
\bibinfo{author}{\bibfnamefont{K.-W.} \bibnamefont{Lee}},
  \bibinfo{author}{\bibfnamefont{J.}~\bibnamefont{Kune\v{s}}},
  \bibnamefont{and} \bibinfo{author}{\bibfnamefont{W.~E.}
  \bibnamefont{Pickett}}, \bibinfo{journal}{Phys. Rev. B}
  \textbf{\bibinfo{volume}{70}}, \bibinfo{pages}{045104}
  (\bibinfo{year}{2004}).

\bibitem[{\citenamefont{Motrunich and Lee}(2004)}]{Motrunich04}
\bibinfo{author}{\bibfnamefont{O.~I.} \bibnamefont{Motrunich}}
  \bibnamefont{and} \bibinfo{author}{\bibfnamefont{P.~A.} \bibnamefont{Lee}},
  \bibinfo{journal}{Phys.\ Rev.\ Lett} \textbf{\bibinfo{volume}{69}},
  \bibinfo{pages}{214516} (\bibinfo{year}{2004}).

\bibitem[{\citenamefont{Mukhamedshin et~al.}(2004)\citenamefont{Mukhamedshin,
  Alloul, Collin, and Blanchard}}]{Mukhamedshin04}
\bibinfo{author}{\bibfnamefont{I.~R.} \bibnamefont{Mukhamedshin}},
  \bibinfo{author}{\bibfnamefont{H.}~\bibnamefont{Alloul}},
  \bibinfo{author}{\bibfnamefont{G.}~\bibnamefont{Collin}}, \bibnamefont{and}
  \bibinfo{author}{\bibfnamefont{N.}~\bibnamefont{Blanchard}},
  \bibinfo{journal}{Phys.\ Rev.\ Lett} \textbf{\bibinfo{volume}{93}},
  \bibinfo{pages}{167601} (\bibinfo{year}{2004}).

\bibitem[{\citenamefont{Shi et~al.}(2003)\citenamefont{Shi, Li, Yu, Zhou,
  Zhang, and Dong}}]{Shi03}
\bibinfo{author}{\bibfnamefont{Y.~G.} \bibnamefont{Shi}},
  \bibinfo{author}{\bibfnamefont{J.~Q.} \bibnamefont{Li}},
  \bibinfo{author}{\bibfnamefont{H.~C.} \bibnamefont{Yu}},
  \bibinfo{author}{\bibfnamefont{Y.~Q.} \bibnamefont{Zhou}},
  \bibinfo{author}{\bibfnamefont{H.~R.} \bibnamefont{Zhang}}, \bibnamefont{and}
  \bibinfo{author}{\bibfnamefont{C.}~\bibnamefont{Dong}},
  \bibinfo{journal}{cond-mat/0306070}  (\bibinfo{year}{2003}).

\bibitem[{\citenamefont{Huang et~al.}(2004{\natexlab{b}})\citenamefont{Huang,
  Khaykovich, Chou, Cho, Lynn, and Lee}}]{huang1}
\bibinfo{author}{\bibfnamefont{Q.}~\bibnamefont{Huang}},
  \bibinfo{author}{\bibfnamefont{B.}~\bibnamefont{Khaykovich}},
  \bibinfo{author}{\bibfnamefont{F.~C.} \bibnamefont{Chou}},
  \bibinfo{author}{\bibfnamefont{J.~H.} \bibnamefont{Cho}},
  \bibinfo{author}{\bibfnamefont{J.~W.} \bibnamefont{Lynn}}, \bibnamefont{and}
  \bibinfo{author}{\bibfnamefont{Y.~S.} \bibnamefont{Lee}},
  \bibinfo{journal}{Phys. Rev. B} \textbf{\bibinfo{volume}{70}},
  \bibinfo{pages}{134115} (\bibinfo{year}{2004}{\natexlab{b}}).

\end{thebibliography}

\end{document}